\documentstyle[12pt,epsf]{article} 

\newcommand{\be}{\begin{equation}}
\newcommand{\ee}{\end{equation}}
\newcommand{\bqn}{\begin{eqnarray}}
\newcommand{\eqn}{\end{eqnarray}}

\begin{document}


\title{Dressed (Renormalized) Coordinates in a Nonlinear System} 
\author{ G. Flores-Hidalgo$^a$\thanks{E-mail: gflores@ift.unesp.br} ~and    
Y. W. Milla$^b$\thanks{E-mail: yonym@cbpf.br}\\ \\
{\it $^a$Instituto de F\'{\i}sica Teorica-IFT/UNESP,}\\
{\it Rua Pamplona 145, 01405-900, S\~ao Paulo, SP, Brazil}
\\
{\it $^b$Centro Brasileiro de Pesquisas Fisicas,}\\
{\it Rua Dr. Xavier Sigaud 150, 22290-180, Rio de Janeiro, RJ, Brazil}
 }
\date{\today} 
 
\maketitle 
 
\begin{abstract}
In previous publications dressed coordinates and
dressed states has been introduced in linear systems. 
Specifically, a system
composed by a harmonic oscillator interacting linearly 
with an infinity set of other oscillators has been treated. 
In this paper we show how to generalize such dressed 
coordinates and dressed states to a nonlinear version of this system.
Also we clarify some misunderstandings about the concept of dressed
coordinates. Indeed now we prefer to call them renormalized
coordinates to emphasize the analogy with the renormalized fields
in quantum field theory. 

\vspace{0.34cm}
\noindent
PACS Number(s):~03.65.Ca, 32.80.Pj

\end{abstract} 
\vskip2pc

\section{Introduction}
In recent publications it has been introduced the concept of dressed coordinates
and dressed states\cite{adolfo1,adolfo2,gabriel,casana}, in 
the context of a
harmonic oscillator (the atom) interacting linearly with a massless scalar 
field. This allowed the construction of dressed states, with the physically 
correct property of stability of the oscillator ground state in the absence of 
field quanta (the quantum vacuum). For a review see the next section.
Also this formalism allowed a nonperturbative 
treatment of the oscillator radiation process. When applied to a confined atom 
(approximated by the oscillator) in a spherical cavity of sufficiently 
small diameter the method accounts for the experimentally observed inhibition 
of the decaying processes \cite{hulet,haroche2}. 

In Ref. \cite{nonlinear} an attempt to construct dressed coordinates and dressed 
states for a nonlinear system has been done. However the approach used there was
more intuitive than formal. The purpose of this paper is to develop a formal
method to construct dressed coordinates in nonlinear systems. 
We will do this by a perturbative expansion in the nonlinear coupling
constant.  To be specific we consider the model with Hamiltonian given by,

\bqn
H&=&\frac{1}{2}\left(p_0^2+\omega_B^2q_0^2\right)+
\frac{1}{2}\sum_{k=1}^N\left(p_k^2+\omega_k^2q_k^2-2c_kq_kq_0\right)\nonumber\\
& &+\sum_{r=0}^N\lambda_{r}{\cal T}^{(r)}_{\mu\nu\rho\sigma}
q_{\mu}q_{\nu}q_{\rho}q_{\sigma}+
\sum_{r=0}^N\alpha_r{\cal R}^{(r)}_{\mu\nu\rho\sigma\tau\epsilon}
q_{\mu}q_{\nu}q_{\rho}q_{\sigma}q_{\tau}q_{\epsilon}\;,
\label{e1}
\eqn
where sums over repeated indices and the limit $N\to\infty$ are understood.
In Eq. (\ref{e1}) the bare frequency of the oscillator, $\omega_B$, is related 
to the physical frequency, $\omega_0$, by \cite{tirring,weiss},

\be
\omega_B^2= \omega_0^2+\sum_{k=1}^N\frac{c_k^2}{\omega_k^2}\;.
\label{e2}
\ee
The coefficients ${\cal T}^{(r)}_{\mu\nu\rho\sigma}$ 
and ${\cal R}^{(r)}_{\mu\nu\rho\sigma\tau\epsilon}$ are
chosen in such a way that the Hamiltonian given by
Eq. (\ref{e1}) is positive defined.

In Ref. \cite{nonlinear}
the quartic nonlinear model, $\alpha_r=0$, has been treated. Here also
we will be mainly interested in the quartic nonlinear model 
to compare with the early treatment.  The sextic nonlinear interaction will
be considered mainly because for some values of $\lambda_r$ and $\alpha_r$ it
is possible to find the exact solution for the ground state and, as explained at
the end of section III, this fact will permit an exact construction of the
dressed coordinates. Then, this sextic model will allow us to test the 
validity of the strategy developed to obtain the dressed coordinates in a general
nonlinear system.

Along this paper we use natural units $c=\hbar=1$.

\section{Defining dressed (renormalized) coordinates and dressed states}
The purpose od this section is twofold. 
First, to make this paper self contained we review what was called dressed
coordinates and dressed states in the Refs. \cite{adolfo1,adolfo2,gabriel}. 
Second, we clarify some misunderstandings about the concept of dressed coordinates,
as we explain below. Indeed now we prefer to call them {\it renormalized} coordinates
to emphasize that these coordinates are analogous to the renormalized fields
in quantum field theory.

To understand the necessity of introducing  dressed coordinates and dressed 
states let us
consider the following problem. Imagine that the oscillator with index zero
in Eq. (\ref{e1}) represents an atom and the other oscillators represent
the modes of the electromagnetic field. If there are no interaction among
them the free Hamiltonian ($c_k=\lambda_r=\alpha_r=0$) has the following 
eigenfunctions, 

\bqn
\psi_{n_0n_1...n_N}(q)&\equiv&
\langle q|n_0,n_1,...,n_N\rangle
\nonumber\\
&=&\prod_{\mu=0}^N\psi_{n_\mu}(q_\mu)\;,
\label{e8a}
\eqn
where $|q\rangle=|q_0,q_1,...,q_N\rangle$ and
$\psi_{n_\mu}(q_\mu)$ is the eigenfunction of a harmonic oscillator
of frequency $\omega_\mu$,

\be
\psi_{n_\mu}(q_\mu)=\left(\frac{\omega_\mu}{\pi}\right)^{1/4}
\frac{H_{n_{\mu}}(\sqrt{\omega_{\mu}}q_\mu)}{\sqrt{2^{n_\mu}n_\mu!}}
e^{-\frac{1}{2}\omega_\mu q_\mu^2}\;.
\label{xe8c} 
\ee
The physical meaning of $\psi_{n_0n_1...n_N}(q)$ in this
case is clear, it represents the atom in its $n_0$-th excited
level and $n_k$ photons of frequencies $\omega_k$. Now, consider the state
of no photons, $\psi_{n_00...0}(q)$: the excited atom in the quantum vacuum. 
We know from experience that any excited level of the atom is unstable,
that is, the state $\psi_{n_00...0}(q)$ is unstable. The explanation of this
fact is that the atom is not isolated from interacting with the quantum 
electromagnetic field, or in other words it is rendered unstable by interacting 
with the quantum vacuum. This interaction in our model is given by
the linear and nonlinear couplings of $q_0$ with $q_k$. Obviously, when we
take into account these interactions any state of the type $\psi_{n_00...0}(q)$ 
will be unstable, since these states are not eigenfunctions of the total
interacting Hamiltonian. But, there is a problem, the state 
$\psi_{00...0}(q)$, that represents the atom in its ground state and no
photons, is also unstable contradicting the experimental fact of the stability
of the ground state in the absence of photons. What is wrong in all this? 
The first thing that cames in our
mind is to think that the model given by Eq. (\ref{e1}) is wrong.
Certainly, we know that the correct theory to describe this physical system is
quantum electrodynamics. On the other hand such a description of this system
could be extremely complicated. If we aim to  maintain the model as simple as 
possible and still insist in describing it by the Hamiltonian given in Eq. (\ref{e1})
what we can do in order to take into account the stability of the ground
state? The answer lies in the spirit of the renormalization program in quantum
fiel theory: the coordinates $q_\mu$ that appear in the Hamiltonian given by 
Eq. (\ref{e1}) are not the physical ones, they are bare coordinates. 
We introduce renormalized coordinates, $q_0'$ and $q_k'$,
respectively for the dressed atom and the dressed photons. We define
these coordinates as the physically meaningful ones. These renormalized 
coordinates were called in prededing works as dressed coordinates 
\cite{adolfo1,adolfo2,gabriel}, for this
reason, from now on we will take these denominations as synonymous. In terms
of these dressed coordinates we define dressed states as

\bqn
\psi_{n_0n_1...n_N}(q')&\equiv&
\langle q'|n_0,n_1,...,n_N\rangle_d \nonumber\\
&=&\prod_{\mu=0}^N\psi_{n_\mu}(q_\mu')\;,
\label{e8b}
\eqn
where the subscript $d$ means dressed state, 
$|q'\rangle=|q_0',q_1',...,q_N'\rangle$ and $\psi_{n_\mu}(q_\mu')$ is given
by

\be
\psi_{n_\mu}(q_\mu')=\left(\frac{\omega_\mu}{\pi}\right)^{1/4}
\frac{H_{n_{\mu}}(\sqrt{\omega_{\mu}}q_\mu')}{\sqrt{2^{n_\mu}n_\mu!}}
e^{-\frac{1}{2}\omega_\mu(q_\mu')^2}\;.
\label{e8c} 
\ee
The dressed states given by Eq. (\ref{e8b}) are defined as the physically measurable
states and describe in general, the physical atom in the
$n_0$-th excited level and $n_k$ physical photons of frequency $\omega_k$.
Obviously, in the limit in which the coupling constants $c_k$, $\lambda_r$ and
$\alpha_r$ 
vanish the renormalized coordinates $q_\mu'$ approach the bare coordinates $q_\mu$. 
Now, in order to relate the bare and dressed coordinates we have to use the physical
requirement of stability of the dressed ground state. The dressed ground
state will be stable only and only if it is defined as the eigenfunction of the
interacting Hamiltonian given in Eq. (\ref{e1}). Also the dressed ground state
must be the one of minimum energy, that is, it must be defined as being
identical (or proportional) to the ground state eigenfunction of the 
interacting Hamiltonian. From this definition, one can construct the dressed
coordinates in terms of the bare ones.

Firstly we explicitly construct the dressed coordinates for the linear
model obtained from Eq. (\ref{e1}) by setting $\lambda_r=\alpha_r=0$,

\be
H_{linear}=\frac{1}{2}\left(p_0^2+\omega_B^2q_0^2\right)+
\frac{1}{2}\sum_{k=1}^N\left(p_k^2+\omega_k^2q_k^2-2c_kq_kq_0\right)\;.
\label{hli}
\ee
Although the task of constructing dressed coordinates in linear systems
has been done in preceding works, we repeat here the calculation in order to 
make this paper self contained. In the next section we will consider the nonlinear 
case. As is well know, the Hamiltonian (\ref{hli}) can be diagonalized by means
of the introduction of normal coordinates $Q_r$ and momenta $P_r$, defined as

\be
q_{\mu}=\sum_{r=0}^Nt_{\mu}^{r}Q_{r}\;,~~~p_{\mu}=\sum_{r=0}^Nt_{\mu}^{r}P_{r}\;,
~~~ \mu=(0,k)\;,~~~k=1,2,...,N\;,
\label{e4}
\ee
where $\{t_{\mu}^{r}\}$ is an orthogonal matrix whose elements are given by
\cite{oconnell,gabrielrudnei}

\be
t_{0}^{r}= \left[1+\sum_{k=1}^{N}\frac{c_{k}^{2}}
{(\omega_{k}^{2}-\Omega_{r}^{2})^{2}}\right]^{-\frac{1}{2}}\;,
~~~~~~~~~~~
t_{k}^{r}=\frac{c_{k}}{(\omega_{k}^{2}-\Omega_{r}^{2})}t_{0}^{r}\;.
\label{e6}
\ee
In normal coordinates the Hamiltonian (\ref{hli}) reads

\be
H_{linear}=\frac{1}{2}\sum_{r=0}^{N}(P_{r}^{2}+\Omega_{r}^{2}Q_{r}^{2})\;,
\label{e5}
\ee
where the $\Omega_{r}$'s are the normal frequencies, corresponding to the 
collective modes and given as solutions of \cite{oconnell,gabrielrudnei},

\be
\omega_0^2-\Omega_r^2=\sum_{k=1}^N\frac{c_k^2\Omega_r^2}
{\omega_k^2(\omega_k^2-\Omega_r^2)}\;.
\label{e7}
\ee
The eigenfunctions of the Hamiltonian given in
Eq. (\ref{e5}) are given by

\bqn 
\phi_{n_{0}n_{1}...n_N}(Q)&\equiv &
\langle Q|n_{0},n_{1},...,n_N\rangle_c
\nonumber\\
&=&\prod_{r=0}^N\phi_{n_r}(Q_r)\;,
\label{ec1}
\eqn
where the subscript $c$ means collective state, $|Q\rangle=|Q_0,Q_1,...,Q_N\rangle$
and $\phi_{n_r}(Q_r)$ are the wave functions corresponding to one dimensional
harmonic oscillators of frequencies $\Omega_r$,

\be
\phi_{n_r}(Q_r)=\left(\frac{\Omega_r}{\pi}\right)^{1/4}
\frac{H_{n_{r}}(\sqrt{\Omega_{r}}Q_r)}{\sqrt{2^{n_r}n_r!}}
e^{-\frac{1}{2}\Omega_rQ_r^2}\;.
\label{e8} 
\ee

Now, the dressed coordinates are defineed  requiring  
$\psi_{00...0}(q')\propto \phi_{00...0}(Q)$, since in this way we guarantee
that $\psi_{00...0}(q')$ is the ground state of $H_{linear}$.  
Then, from Eqs. (\ref{ec1}) and  (\ref{e8b}), we have

\be
e^{-\frac{1}{2}\sum_{\mu=0}^N\omega_\mu (q_\mu')^2}
\propto e^{-\frac{1}{2}\sum_{r=0}^N\Omega_r Q_r^2}\;,
\label{aux1}
\ee
from which we obtain 
\be
q_\mu'=\sum_{r=0}^N\sqrt{\frac{\Omega_r}{\omega_\mu}}t_\mu^rQ_r\;,
\label{e10}
\ee
as can be seen by direct substitution in Eq. (\ref{aux1}) and using the
orthonormality properties of the $\{t_\mu^r\}$ matrix.
The above definition guarantees the stability of the dressed ground state,
however, since the other dressed states are not energy eigenfunctions, 
they will not remain stable. For example the first excited dressed state,
whose eigenfunction is $\psi_{10...0}(q')$, will decay to the ground state
$\psi_{00...0}(q')$.

We have to remark here that the dressed coordinates here introduced are not 
simply a change of variables, they are new coordinates in its own right and
are introduced by physical consistence requirement of the model. Also we have 
to stress that our dressed coordinates are not the same as the ones employed in
other references, as
for example in \cite{prigogine} and references therein, where the authors called 
dressed coordinates the collective normal ones. Also our dressed states
are very different from the ones defined in Refs. 
\cite{polonsky,haroche,cohen,cohen1}, where the authors called dressed
states the states obtained by diagonalizing a truncated finite matrix
representation of the Hamiltonian.

Before leaving this section it will be useful to establish the relation
between $\psi_{n_0n_1...n_N}(q')=\langle q'|n_0,n_1,...,n_N\rangle_d$ and 
$\langle Q|n_0,n_1,...,n_N\rangle_d$. For this end we write

\bqn
_d\langle n_0,n_1,...,n_N|m_0,m_1,...,m_N\rangle_d&=&
 \int dq'~_d\langle n_0,n_1,...,n_N|q'\rangle\langle q'|m_0,m_1,...,m_N\rangle\nonumber\\
&=&\int dQ \left|\frac{\partial q'}{\partial Q}\right|
~_d\langle n_0,n_1,...,n_N|q'\rangle\langle q'|m_0,m_1,...,m_N\rangle_d\nonumber\\
&=&\int dQ~_d\langle n_0,n_1,...,n_N|Q\rangle\langle Q|m_0,m_1,...,m_N\rangle_d\;,
\label{aux2}
\eqn
where $dq'=\prod_{\mu=0}^Ndq_\mu'$, $dQ=\prod_{r=0}^NdQ_r$ and 
$\left|\partial q'/\partial Q\right|$ is the Jacobian associated
to the transformation $q'_\mu\to Q_r$. From Eq. (\ref{aux2}) we get

\be
\langle Q|n_0,n_1,...,n_N\rangle_d=\left|\frac{\partial q'}{\partial Q}\right|^{1/2}
\langle q'|n_0,n_1,...,n_N\rangle_d\;.
\label{aux3}
\ee
Taking $n_0=n_1=...=n_N=0$ in Eq. (\ref{aux3}) and using 
$\psi_{00...0}(q')\propto\phi_{00...0}(Q)$ we get

\be
|0,0,...,0\rangle_d\propto\int dQ\left|\frac{\partial q'}{\partial Q}\right|^{1/2}
|Q\rangle\langle Q|0,0,...,0\rangle_c  \;.
\label{aux4}
\ee
In the linear case, we easily get, from Eq. (\ref{e10}),
$\left|\partial q'/\partial Q\right|=
\prod_{r,\mu=0}^N\sqrt{\Omega_r/\omega_\mu}$ and using this result in
Eq. (\ref{aux4}) we obtain

\be
|0,0,...,0\rangle_d\propto
|0,0,...,0\rangle_c\;.
\label{aux5}
\ee
For a nonlinear system, certainly a relation of the type given by Eq. 
(\ref{aux5}) will not hold. 

In next section we construct dressed coordinates in the
nonlinear model described by the Hamiltonian given in Eq. (\ref{e1}).

\section{Constructing renormalized coordinates in a nonlinear model}
Now we are ready to construct dressed coordinates and dressed states in
the nonlinear model given by Eq. (\ref{e1}). For this purpose we have
to find, firstly, the eigenfunctions of this Hamiltonian, in particular
its ground state eigenfunction.

In order to maintain things
as simple as possible and to compare with the the preceding treatment
given in Ref. \cite{nonlinear}, we consider the nonlinear quartic interaction
obtained from the model given described in Eq. (\ref{e1}) by setting $\alpha_r=0$.
Following Ref. \cite{nonlinear} we make the simplest choice for the coefficients 
${\cal T}^{(r)}_{\mu\nu\rho\sigma}$ as 

\be
{\cal T}^{(r)}_{\mu\nu\rho\sigma}=t_\mu^rt_\nu^rt_\rho^rt_\sigma^r\;.
\label{e11}
\ee
Substituting Eqs. (\ref{e4}) and (\ref{e11}) in Eq. (\ref{e1}) we get

\be
H=\frac{1}{2}\sum_{r=0}^N\left(P_r^2+\Omega_r^2Q_r^2+2\lambda_rQ_r^4\right)\;,
\label{e12}
\ee
that is, we obtain a system of uncoupled quartic anharmonic oscillators.
In Eq. (\ref{e12})
we can notice that $\lambda_r$ has dimension of $[frequency]^3$. Then
we write $\lambda_r=\lambda\Omega_r^3$, where $\lambda$ is a dimensionless
coupling constant. The eigenfunctions of the Hamiltonian given by Eq. (\ref{e12})
can be written as

\bqn
\phi_{n_0n_1...n_N}(Q;\lambda) &\equiv&
\langle Q|n_0,n_1,...,n_N;\lambda\rangle_c
\nonumber\\
&=&\prod_{r=0}^N\phi_{n_r}(Q_r;\lambda)\;,
\label{e13}
\eqn
where $\phi_{n_r}(Q_r;\lambda)$ are eigenfunctions of 
$\left(P_r^2+\Omega_r^2Q_r^2+2\lambda\Omega_r^3Q_r^4\right)/2$ and can be 
written formally as (see Appendix) 

\be
\phi_{n_r}(Q_r;\lambda)=\left(\frac{\Omega_r}{\pi}\right)^{1/4}
\left[\frac{H_{n_r}(\sqrt{\Omega_r}Q_r)}{\sqrt{2^{n_r}n_r!}}+
\sum_{l=1}^\infty\lambda^lG_{n_r}^{(l)}(\sqrt{\Omega_r}Q_r)\right]
e^{-\frac{\Omega_r}{2}Q_r^2}\;,
\label{ec4}
\ee
where $G_{n_r}^{(l)}(\sqrt{\Omega_r}Q_r)$  are linear combinations of Hermite 
polynomials. The corresponding eigenvalues for the Hamiltonian given in 
Eq. (\ref{e12}) are given by,

\be
E_{n_0n_1...n_N}(\lambda)=\sum_{r=0}^NE_{n_r}(\lambda)\;,
\label{e13a}
\ee
where $E_{n_r}(\lambda)$ are the eigenvalues corresponding to the eigenstates
given in Eq. (\ref{ec4}), 

\be
E_{n_r}(\lambda)=(n_r+\frac{1}{2})\Omega_r+
\sum_{l=1}^\infty \lambda^l E^{(l)}_{n_r}\;,
\label{e13b}
\ee
with the $E^{(l)}_{n_r}$ obtained by using standard perturbation theory (see Appendix).

Taking $n_0=n_1=...=n_N=0$ in Eq. (\ref{e13}) we get for the ground state 
eigenfunction of the total system,

\be
\phi_{00...0}(Q;\lambda)=
\prod_{r=0}^N\left(\frac{\Omega_r}{\pi}\right)^{1/4}
\left[1+\sum_{l=1}^{\infty}\lambda^lG_{0}^{(l)}(\sqrt{\Omega_r}Q_r)\right]
e^{-\frac{\Omega_r}{2}Q_r^2}\;.
\label{e14}
\ee
To properly define (see comments below)
the dressed coordinates it is convenient to write the
above equation as,

\bqn
\phi_{00...0}(Q;\lambda)&=&
\prod_{r=0}^N\left(\frac{\Omega_r}{\pi}\right)^{1/4}
\left[1+\sum_{l=1}^{\infty}\lambda^lG_{0}^{(l)}(0)+
\sum_{l=1}^{\infty}\lambda^l\left(G_{0}^{(l)}(\sqrt{\Omega_r}Q_r)
-G_{0}^{(l)}(0)\right)\right]e^{-\frac{\Omega_r}{2}Q_r^2}\nonumber\\
&\propto&
\prod_{r=0}^N\left[1+\sum_{n=0}^\infty(-1)^n\right.
\sum_{l_0l_1...l_n=1}^{\infty}
\lambda^{l_0+l_1+...+l_n}\nonumber\\
& &~~~~~~~~~~~\times\left.
\left(G_{0}^{(l_0)}(\sqrt{\Omega_r}Q_r)-G_{0}^{(l_0)}(0)\right)
G_0^{(l_1)}(0)...G_0^{(l_n)}(0) \right]e^{-\frac{\Omega_r}{2}Q_r^2}\;,
\label{e14b}
\eqn
where in the second line we factored the term 
$1+\sum_{l=1}^{\infty}\lambda^lG_{0}^{(l)}(0)$ and used 
$(1+x)^{-1}=\sum_{n=0}^\infty(-1)^nx^n$.

The physically measurable states, the dressed states, are defined by
Eqs. (\ref{e8b}) and (\ref{e8c}). Hence, the dressed 
coordinates $q_\mu'$ will be defined in such a way that the dressed
ground state equals (or is proportional) to the ground state of the
nonlinear interacting Hamiltonian given in Eq. (\ref{e14b}).
That is, we define the dressed coordinates imposing the condition 
$\psi_{00...0}(q')\propto\phi_{00...0}(Q;\lambda)$ which by using Eqs. (\ref{e8b}),
(\ref{e8c}) and (\ref{e14b}) can be written as

\bqn
\!\!\!\!\!\!\!\!\!\!\!
e^{-\frac{1}{2}\sum_{\mu=0}^N\omega_\mu(q_\mu')^2}\!\!\!\!&=&\!\!\!
\prod_{r=0}^N\left[1+
\sum_{n=0}^\infty(-1)^n\right.\sum_{l_0l_1...l_n=1}^{\infty}
\lambda^{l_0+l_1+...+l_n}
\nonumber\\
& &\left.~~~~~~~~~\times
\left(G_{0}^{(l_0)}(\sqrt{\Omega_r}Q_r)-G_{0}^{(l_0)}(0)\right)
G_0^{(l_1)}(0)...G_0^{(l_n)}(0)\right]e^{-\frac{\Omega_r}{2}Q_r^2}\;.
\label{e16}
\eqn
Now, we write a perturbative expansion in $\lambda$ for $q_\mu'$,

\be
q'_\mu=\sum_{r=0}^N\sqrt{\frac{\Omega_r}{\omega_\mu}}t_\mu^r\left[Q_r
+\frac{1}{\sqrt{\Omega_r}}
\sum_{l=1}^{\infty}\lambda^lF_r^{(l)}(\sqrt{\Omega_r}Q_r)\right]\;.
\label{e15}
\ee
Replacing Eq. (\ref{e15}) in Eq. (\ref{e16}) and using the the orthonormality
of the matrix $\{t_\mu^r\}$ we get

\bqn
&&\!\!\!\!\!\!\!\!\exp\left[-\sum_{l=1}^\infty\lambda^l 
\sqrt{\Omega_r}Q_r F_r^{(l)}(\sqrt{\Omega_r}Q_r)
-\frac{1}{2}\sum_{l,m=1}^\infty
\lambda^{l+m}F_r^{(l)}
(\sqrt{\Omega_r}Q_r)F_r^{(m)}(\sqrt{\Omega_r}Q_r)\right]\nonumber\\
&&\!\!\!\!\!\!\!\!\!\!\!\!
=1+\sum_{n=0}^\infty (-1)^n\sum_{l_0l_1...l_n=1}^{\infty}
\lambda^{l_0+l_1+...+l_n}
\left(G_{0}^{(l_0)}(\sqrt{\Omega_r}Q_r)-G_{0}^{(l_0)}(0)\right)
G_0^{(l_1)}(0)...G_0^{(l_n)}(0)
\;.
\label{e16a}
\eqn
Expanding the exponential in the left hand side of Eq. (\ref{e16a}) and
identifying equal powers of $\lambda$ we can obtain all the 
$F^{(l)}_r(\sqrt{\Omega_r}Q)$.
The general expression is very complicated, here we only write the
first three terms:

\be
F_r^{(1)}(\xi_r)=-\frac{1}{\xi_r}
\left(G_0^{(1)}(\xi_r)-G_0^{(1)}(0)\right)\;,
\label{e18}
\ee
\be
F_r^{(2)}(\xi_r)=-\frac{1}{\xi_r}
\left[G^{(2)}_0(\xi_r)-G_0^{(2)}(0)+G_0^{(1)}(0)\xi_rF_r^{(1)}(\xi_r)
+\frac{1}{2}(1-\xi_r^2)\left(F_r^{(1)}(\xi_r)\right)^2\right]
\label{e19}
\ee
and

\bqn
F_r^{(3)}(\xi_r)&=&-\frac{1}{\xi_r}
\left[G^{(3)}_0(\xi_r)-G_0^{(3)}(0)
-\left(G_0^{(2)}(\xi_r)-G_0^{(2)}(0)\right)G_0^{(1)}(0)\right.\nonumber\\
& &~~~~~~~
+\xi_rF_r^{(1)}(\xi_r)\left(G_0^{(2)}(0)-\left(G_0^{(1)}(0)\right)^2\right)
+\frac{\xi_r}{3!}\left(\xi_r^2-3\right)\left(F_r^{(1)}(\xi_r)\right)^3
\nonumber\\
& &~~~~~~~\left.+
(1-\xi_r^2)F_r^{(1)}(\xi_r)F_r^{(2)}(\xi_r)\right]\;,
\label{e20}
\eqn
where $\xi_r=\sqrt{\Omega_r}Q_r$. 
From the Appendix, using Eqs. (\ref{xx1}-\ref{xx3a}) in Eqs. (\ref{e18}) and
(\ref{e19}) we get respectively,

\be
F_r^{(1)}(\xi_r)=\frac{1}{4}\left(3\xi_r+\xi_r^3\right)
\label{e20a}
\ee
and

\be
F_r^{(2)}(\xi_r)=-\frac{1}{16}\left(\frac{93}{2}\xi_r+
14\xi_r^3+\frac{11}{6}\xi_r^5\right)\;.
\label{e20b}
\ee
Replacing above equations in Eq. (\ref{e15}) we obtain at order $\lambda^2$,

\be
\xi_\mu'=\sum_{r=0}^Nt_\mu^r\left[\xi_r+
\frac{\lambda}{4}\left(3\xi_r+\xi_r^3\right)-
\frac{\lambda^2}{16}\left(\frac{93}{2}\xi_r+
14\xi_r^3+\frac{11}{6}\xi_r^5\right)
+{\cal O}(\lambda^3)\right]\;,
\label{e20c}
\ee
where we have introduced the dimensionless dressed coordinate 
$\xi_\mu'=\sqrt{\omega_\mu}q_\mu'$. 

Before leaving this section we would like to comment  why we factored
the term $1+\sum_{l=1}^{\infty}\lambda^lG_{0}^{(l)}(0)$ in Eq. (\ref{e14b}).
Notice that we define the dressed coordinates $q_\mu'$ by means of the 
proportionality $\psi_{00...0}(q')\propto\phi_{00...0}(Q;\lambda)$. To promote 
this proportionality into an equality we have
to take care in defining a well behaved transformation between dressed 
and collective coordinates, for example it would be undesirable any singular
transformation. To see how the above undesirable situation occurs, define
the dressed coordinates through Eq. (\ref{e14}) without the factorization
of the term $1+\sum_{l=1}^{\infty}\lambda^lG_{0}^{(l)}(0)$. It is easy to 
show that in such a case the transformation between $q_\mu'$ and $Q_r$ is
singular. For example we obtain for $F_r^{1}(\xi_r)$ and $F_r^{2}(\xi_r)$ 

\be
F_r^{(1)}(\xi_r)=-\frac{1}{\xi_r}
G_0^{(1)}(\xi_r)\;,
\label{e20d}
\ee
\be
F_r^{(2)}(\xi_r)=-\frac{1}{\xi_r}
\left[G^{(2)}_0(\xi_r)+
\frac{1}{2}(1-\xi_r^2)
\left(F_r^{(1)}(\xi_r)\right)^2\right]
\label{e20f}\;.
\ee
Since $G_0^{(1)}(\xi_r)$ and $G_0^{(2)}(\xi_r)$ are not homogeneus
functions of $\xi_r$ [see Appendix, Eqs. (\ref{xx1}) and (\ref{xx2})] 
then Eqs. (\ref{e20d}) and (\ref{e20f}) are
singular in $\xi_r=0$. Consequently, the dressed coordinates 
defined through this prescription are not well defined. To understand
what is happening and how to remedy this problem, note that this singularity 
means that  $\xi_rF_r^{(l)}(\xi_r)$ is not homegeneus in $\xi_r$. But the
effect of this nonhomogeneous term on the wave function [that contains
terms of the type $e^{-\lambda^l\xi_rF_r^{(l)}(\xi_r)}$, see Eq. (\ref{e16a})]
is just equal to a factorization term. Then to remedy the situation we have 
to make a  convenient factorization in $\psi_{00..0}(q')$ or, equivalently, in
$\phi_{00..0}(Q)$
before promoting the proportionality into an equality. That our choice,
the factorization of $1+\sum_{l=1}^{\infty}\lambda^lG_{0}^{(l)}(0)$
in Eq. (\ref{e14b}), is the correct one is supported by the fact that we obtain
well behaved dressed coordinates. To further support our choice,
we consider a system in which we can solve exactly for the ground state,
allowing us to obtain exact dressed coordinates. Comparing these exact
dressed coordinates and the perturbative ones we get the same answer.
The model in mention is the one whose Hamiltonian is given by Eq. (\ref{e1})
with coupling constants defined as

\be
\lambda_r{\cal T}^{(r)}_{\mu\nu\rho\sigma}=
\frac{\lambda\Omega_r^3}{(1-3\lambda)^{3/2}}
t_\mu^r t_\nu^r t_\rho^r t_\sigma^r,~~~
\alpha_r{\cal R}^{(r)}_{\mu\nu\rho\sigma\tau\epsilon}=
\frac{\lambda^2\Omega_r^4}{2(1-3\lambda)^2}
t_\mu^r t_\nu^r t_\rho^r t_\sigma^r t_\tau^r t_\epsilon^r\;.
\label{e20g}
\ee
Using the above expression in Eq. (\ref{e1}) we get a system of uncoupled
sextic anharmonic oscillators,

\be
H=\frac{1}{2}\sum_{r=0}^N\left(P_r^2+\Omega_r^2Q_r^2+
\frac{2\lambda\Omega_r^3}{(1-3\lambda)^{3/2}}Q_r^4+
\frac{\lambda^2\Omega_r^4}{(1-3\lambda)^2}Q_r^6\right)\;.
\label{e20h}
\ee
By direct substitution it is easy to show that the above Hamiltonian
have as ground state eigenfunction the following solution \cite{skala}

\be
\phi(Q;\lambda)={\cal N}
e^{-\sum_{r=0}^N\left(\beta_rQ_r^2+\lambda\beta_r^2Q_r^4\right)}\;,
\label{e20i}
\ee
where ${\cal N}$ is a normalization constant,

\be
\beta_r=\frac{\Omega_r}{2\sqrt{1-3\lambda}}
\label{ccc}
\ee
and the corresponding ground state energy is given by 

\be
E(\lambda)=\sum_{r=0}^N\beta_r\;.
\label{exact}
\ee
Now, the dressed coordinates can be defined by 

\be
e^{-\sum_{\mu=0}^N\omega_\mu (q_\mu')^2}=
e^{-\sum_{r=0}^N\left(\beta_rQ_r^2+\lambda\beta_r^2Q_r^4\right)}\;,
\label{e20j}
\ee
from which we obtain

\be
\xi_\mu'=\sum_{r=0}^Nt_\mu^r\xi_r\left(\frac{1}{\sqrt{1-3\lambda}}
+\frac{\lambda\xi_r^2}{2(1-3\lambda)}\right)^{1/2}\;.
\label{e20k}
\ee
Note that at order $\lambda$ both the quartic and sextic anharmonic
Hamiltonians, given respectively by Eqs. (\ref{e12}) and (\ref{e20h}),
are equivalent. Then, if our strategy to define perturbatively
the dressed coordinates is correct, at order $\lambda$ Eq. (\ref{e20c})
must be identical to Eq. (\ref{e20k}). Expanding Eq. (\ref{e20k}) at order 
$\lambda$ we can see that it is indeed the case. Then we conclude that
our strategy for defining perturbatively the dressed coordinates is
the correct one.

\section{The decay process of the first excited state}
In Ref. \cite{nonlinear} the probability of the particle oscillator to remain
in the first excited state has been computed at first order for the nonlinear
quartic interaction. However, as we have already mentioned, the approach used
there was more intuitive than formal. In order to see in what extent such
calculation is correct, in this section we compute the same quantity by
using the formalism presented in the last section.
To maintain the reasoning as general as possible we present the steps necessary 
to compute the probability amplitude associated with the most general transition,

\be
{\cal A}_{n_0n_1...n_N}^{m_0m_1...m_N}(t)=~
_d\langle m_0,m_1,...,m_N|e^{-iHt}|n_0,n_1,...,n_N\rangle_d\;,
\label{e21}
\ee
that is, we prepare our system initially at time $t=0$ in the dressed
state $|n_0,n_1,...,n_N\rangle_d$, then we ask what is the  probability amplitude 
of finding, in a measurement performed at time $t$, the dressed state 
$|m_0,m_1,...,m_N\rangle_d$.  Introducing a complete set of eigenstates of
the total Hamiltonian $H$, given by Eq. (\ref{e13}), in Eq. (\ref{e21}) we find  

\bqn
{\cal A}_{n_0n_1...n_N}^{m_0m_1...m_N}(t)\!\!\!\!&=&\!\!\!\!
\sum_{l_0l_1...l_N=0}^\infty\!\!\!~_d\langle m_0,\!m_1,...,\!m_N|e^{-iHt}
|l_0,\!l_1,...,\!l_N;\!\lambda\rangle_c~\!
_c\langle l_0,\!l_1,...,\!l_N;\!\lambda|n_0,\!n_1,...,\!n_N\rangle_d
\nonumber\\
&=&\sum_{l_0l_1...l_N=0}^\infty T^{l_0l_1...l_N}_{n_0n_1...n_N}(\lambda)
T^{l_0l_1...l_N}_{m_0m_1...m_N}(\lambda)
e^{-itE_{l_0l_1...l_N}(\lambda)}\;,
\label{e22}
\eqn
with

\bqn
T^{l_0l_1...l_N}_{n_0n_1...n_N}(\lambda)&=&\int dQ~
_c\langle l_0,l_1,...,l_N;\lambda|Q\rangle \langle Q|
n_0,n_1,...,n_N\rangle_d\nonumber\\
&=&\int dQ \left|\frac{\partial q'}{\partial Q}\right|^{1/2}
\phi_{l_0l_1...l_N}(Q;\lambda)\psi_{n_0n_1...n_N}(q')\;,
\label{e23}
\eqn
where in the second line we have used Eq. (\ref{aux3}).

From Eq. (\ref{e15}) we get easily the Jacobian $|\partial q'/\partial Q|$,

\be
\left|\frac{\partial q'}{\partial Q}\right|=\prod_{r,\mu=0}^N
\left|\sqrt{\frac{\Omega_r}{\omega_\mu}}
\left(1+\frac{1}{\sqrt{\Omega_r}}\sum_{l=1}^\infty\lambda^l
\frac{\partial}{\partial Q_r}F_r^{(l)}(\sqrt{\Omega_r}Q_r)\right)\right|\;.
\label{e23b}
\ee

Now we evaluate, at first order in $\lambda$, the probability amplitude
to the particle oscillator remain at time $t$ in the first excited
state if it has been prepared in that state at time $t=0$. This quantity
is obtained taking $n_0=m_0=1$ and $n_k=m_k=0$ in Eq. (\ref{e22}).
Notice that to compute ${\cal A}_{10...0}^{10...0}(t)$ at first order
in $\lambda$ we need to find 
$T^{l_0l_1...l_N}_{10...0}(\lambda)$, defined in Eq. (\ref{e23}), 
at order $\lambda$.
Replacing Eq. (\ref{e20a}) in Eq. (\ref{e23b}) we get, 

\be
\left|\frac{\partial q'}{\partial Q}\right|=
\left(\prod_{\mu,r=0}^N\frac{\Omega_r}{\omega_\mu}\right)^{1/4}
\left[1+\frac{3\lambda}{32}\sum_{s=0}^N\left(6H_0(\sqrt{\Omega_s}Q_s)
+H_2(\sqrt{\Omega_s}Q_s\right)\right]+{\cal O}(\lambda^2)\;.
\label{ab23}
\ee
At order $\lambda$, from Eq. (\ref{e13}), we have for 
$\phi_{l_0l_1...l_N}(Q,\lambda)$,

\be
\phi_{l_0l_1...l_N}(Q,\lambda)=\prod_{r=0}^N\phi_{l_r}(Q_r)
+\lambda\sum_{r=0}^N\left[\left(\frac{\Omega_r}{\pi}\right)^{1/4}
G_{l_r}^{(1)}(\sqrt{\Omega_r}Q_r)e^{-\frac{\Omega_r}{2}Q_r^2}\prod_{s\neq r}
\phi_{l_s}(Q_s)\right]+{\cal O}(\lambda^2)\;,
\label{ab24}
\ee
where the $\phi_{l_r}(Q_r)$ are given by Eq. (\ref{e8}),

\bqn
G_{l_r}^{(1)}(\sqrt{\Omega_r}Q_r)&=&a_{l_r}H_{l_r-4}(\sqrt{\Omega_r}Q_r)+
b_{l_r}H_{l_r-2}(\sqrt{\Omega_r}Q_r)
+c_{l_r}H_{l_r+2}(\sqrt{\Omega_r}Q_r)\nonumber\\
& &
+d_{l_r}H_{l_r+4}(\sqrt{\Omega_r}Q_r)\;,
\label{ab25}
\eqn
and $a_{l_r},~b_{l_r},~c_{l_r}$ and $d_{l_r}$ are given in Appendix, Eq. (\ref{xx3}). 
Using Eq. (\ref{e16}) and Eq. (\ref{e15}) we have for $\psi_{10...0}(q')$, 

\bqn
\!\!\!\!\!\!\!\!
\psi_{10...0}(q')&=&\left(\prod_{\mu=0}^N\frac{\omega_\mu}{\pi}\right)^{1/4}
\frac{H_1(\sqrt{\omega_0}q_0')}{\sqrt{2}}
e^{-\frac{1}{2}\sum_{\mu=0}^N\omega_\mu(q_\mu')^2}\nonumber\\
&=&\left(\prod_{\mu=0}^N\frac{\omega_\mu}{\pi}\right)^{1/4}
\sum_{r=0}^N\frac{t_0^r}{\sqrt{2}}\left[H_1(\sqrt{\Omega_r}Q_r)
+2\lambda F_r^{(1)}(\sqrt{\Omega_r}Q_r)
\right.\nonumber\\
& &~~~
-\lambda H_1(\sqrt{\Omega_r}Q_r)\sum_{s=0}^N\sqrt{\Omega_s}Q_s\left.
F_s^{(1)}(\sqrt{\Omega_s}Q_s)\right]
e^{-\frac{1}{2}\sum_{u=0}^N\Omega_uQ_u^2}+{\cal O}(\lambda^2)\;.
\label{ab26}
\eqn
Replacing Eqs. (\ref{ab23}), (\ref{ab24}) and (\ref{ab26}) in Eq. (\ref{e23})
we obtain after a long, but straightforward, calculation

\bqn
T^{l_0l_1...l_N}_{10...0}(\lambda)&=&
\sum_{r=0}^Nt_0^r\delta_{l_r1}\prod_{s\neq r}\delta_{l_s0}
+9\frac{\sqrt{6}}{16}\lambda\sum_{r=0}^Nt_0^r\delta_{l_r3}\prod_{s\neq r}
\delta_{l_s0}\nonumber\\
& &+3\frac{\sqrt{2}}{16}\lambda\sum_{r\neq s}t_0^r\delta_{l_r1}
\delta_{l_s2}\prod_{u\neq r,s}\delta_{l_u0}+{\cal O}(\lambda^2)\;.
\label{e26}
\eqn

Replacing Eq. (\ref{e26}) in Eq. (\ref{e22}) and using from the Appendix
$E_{1_r}(\lambda)\!\approx\!\frac{3}{2}\Omega_r\!+\!\frac{15}{4}\lambda\Omega_r$ 
we obtain for  ${\cal A}_{10...0}^{10...0}(t)$, which we denote as 
$f_{00}(t;\lambda)$,
 
\be
f_{00}(t;\lambda)=e^{-\frac{it}{2}\sum_{r=0}^N\Omega_r}
\sum_{r=0}^N(t_0^r)^2\left(1-\frac{15}{4}i\lambda t\Omega_r\right)
e^{-it\Omega_r}+{\cal O}(\lambda^2)\;.
\label{e27}
\ee
From the above equation we get the probability to the 
particle oscillator remain in the first excited level, 

\be
|f_{00}(t;\lambda)|^2=|f_{00}(t)|^2+\frac{15\lambda t}{4}
\frac{\partial}{\partial t}|f_{00}(t)|^2+
{\cal O}(\lambda^2)\;,
\label{e29}
\ee
where 

\be
f_{00}(t)=\sum_{r=0}^N(t_0^r)^2e^{-i\Omega_r t}. 
\label{e30}
\ee

Equation (\ref{e29}) is the same as the one obtained in Ref.
\cite{nonlinear}. We obtained the same result because at order $\lambda$
the square of $T_{10...0}^{l_0l_1...l_N}$ is given only by the
square of the first term in Eq. (\ref{e26}), that does not depend
on $\lambda$. Then, the effects
of the nonlinearities, at this order, cames only from the 
corrections to the energies, as was assumed in Ref. \cite{nonlinear}.

\section{Conclusions}
In this paper, after clarifying what we understand by dressed coordinates and
dressed states,  we have developed a formal method to construct perturbatively dressed 
coordinates in  nonlinear systems. Although we restricted our calculations to a 
very special quartic interacting  term, we have pointed out the necessity of 
factoring a term in order to avoid artificial singularities which otherwise would appear
if we do not make such factorization. That this factorization is the correct one
has been checked by using an exactly solvable sextic interacting model. Then,
in more general nonlinear systems, one can follow the same procedure to 
construct the dressed coordinates. 

At the end of section II we remarked that for nonlinear systems, in the number 
representation,
the dressed ground state is not equivalent to the ground state of the 
total system, see Eq. (\ref{aux4}). This fact must not be seen as in contradiction with 
our definition of dressed coordinates, since we have defined them by requiring the
equivalence of the dressed ground state in dressed coordinates representation
and the ground state of the system in normal coordinates representation.
 We can understand the mentioned  non equivalence, by noting 
that although  the dressed ground state is an eigenstate of the dressed number 
operators (associated with the dressed coordinates) the ground state of the system, 
in general, is not an eigenstate of the collective (normal) number operators. 
For example,
in the quartic nonlinear  case one can easily verify that the ground state 
(and also, the other eigenstates) is not an eigenstate of the collective number 
operators, but a linear superposition of eigenstates of these operators [see
Appendix, Eq. (\ref{ap5})].

Finally, we considered the computation of the probability of the particle
oscillator to remain excited in its first excited level, and showed that the result
coincides with the result obtained in Ref. \cite{nonlinear}. Then, one of the 
conclusions of Ref. \cite{nonlinear} remains: the effect of the nonlinear quartic terms 
is the enhancement 
of the decay of the particle oscillator from its first excited level
to the ground state. This fact, can be easily seen from Eq. (\ref{e29}) by
noticing that $|f_{00}(t)|^2$ (the probability to the particle oscillator remain
in the first excited state in the absence of nonlinear interactions) 
in free space, is a decreasing  (almost exponentially) function of time.

\vspace{0.5cm}

\begin{center}
{\large \bf Acknowledgements}
\end{center}
We acknowledge A. P. C. Malbouisson (CBPF) for reading the manuscript.
GFH  is supported by FAPESP, grant 02/09951-3 and YWM is 
supported by a grant from CNPq (Conselho Nacional de Desenvolvimento
Cientifico e Tecnol\'ogico).

\appendix
\section*{Appendix: The perturbed eigenfunctions and eigenvalues}
It is easy to see that the eigenfunctions of the quartic anharmonic
oscillator can be written formally as those given in Eq. (\ref{ec4}).
We have to notice only that any wave function can be expanded
in the basis $\phi_{n}(Q)$ (we omit here the index $r$), given by the eigenvalues of the
linear part of the Hamiltonian. And since  $\phi_n(Q)$ are
given by $\exp(-\Omega Q^2/2)$ times a Hermite polynomial of degree $n$,
we see that an expression of the type given in Eq. (\ref{ec4})
follows. In what follows we compute $G_n^{(1)}(\sqrt{\Omega}Q)$ and 
$G_n^{(2)}(\sqrt{\Omega}Q)$ by using standard perturbation theory.

At second order in standard perturbation theory the eigenfunctions
and eigenvalues of a Hamiltonian $\hat{H}=\hat{H}_0+\lambda\hat{V}$ 
are given respectively by,

\be
|n,\lambda\rangle=|n\rangle+\lambda\sum_{k\neq n}\frac{V_{kn}|k\rangle}
{E_n-E_k}+
\lambda^2\left(\sum_{k,l\neq n}\frac{V_{kl}V_{ln}|k\rangle}
{(E_n-E_k)(E_n-E_l)}-V_{nn}\sum_{k\neq n}
\frac{V_{kn}|k\rangle}{(E_n-E_k)^2}\right)+{\cal O}(\lambda^3)
\label{ap1}
\ee
and

\be
E_n(\lambda)=E_n+\lambda V_{nn} +\lambda^2\sum_{k\neq n}\frac{|V_{nk}|^2}
{E_n-E_k}+{\cal O}(\lambda^3)\;,
\label{ap2}
\ee
where 

\be
V_{kn}=\langle k|\hat{V}|n\rangle
\label{ap3}
\ee
and $|n\rangle$ and $E_n$ are respectively eigenfunctions and 
eigenvalues of the unperturbed Hamiltonian $\hat{H}_0$.

For the anharmonic oscillator with $\hat{V}=\Omega^3 \hat{Q}^4$ we
obtain easily

\bqn
V_{kn}&=&\frac{\Omega}{4}[\sqrt{k_4}\delta_{k,n-4}
+2(2n-1)\sqrt{k_2}\delta_{k,n-2}+3(2n^2+2n+1)\delta_{k,n}
\nonumber\\
& &~~~
+2(2n+3)\sqrt{n_2}\delta_{k,n+2}+\sqrt{n_4}\delta_{k,n+4}]\;,
\label{ap4}
\eqn
where $k_n=(k+1)(k+2)...(k+n)$. Replacing 
Eq. (\ref{ap4}) in Eqs. (\ref{ap1}) and (\ref{ap2}) we obtain
respectively,

\bqn
|n,\lambda\rangle &=&|n\rangle +\lambda\left(a_n'|n-4\rangle+b_n'|n-2\rangle
+c_n'|n+2\rangle+d_n'|n+4\rangle\right)\nonumber\\
& &~~~~+\lambda^2\left(e_n'|n-8\rangle+f_n'|n-6\rangle+g_n'|n-4\rangle
+h_n'|n-2\rangle
\right.\nonumber\\
& &~~~~~~~~~~~
\left.
+t_n'|n+2\rangle+u_n'|n+4\rangle+v_n'|n+6\rangle+w_n'|n+8\rangle\right)+
{\cal O}(\lambda^3)
\label{ap5}
\eqn
and

\be
E_n(\lambda)=(n+\frac{1}{2})\Omega+\lambda E_n^{(1)}
+\lambda^2 E_n^{(2)}+{\cal O}(\lambda^3)\;,
\label{ap6}
\ee
where

\bqn
a_n'&=&\frac{1}{16}\sqrt{(n-4)_4}\;,~~~
b_n'=\frac{(2n-1)}{4}\sqrt{(n-2)_2}\;,\nonumber\\
c_n'&=&-\frac{(2n+3)}{4}\sqrt{n_2}\;,~~~
d_n'=-\frac{1}{16}\sqrt{n_4}\;;
\label{xx3ap}
\eqn
\bqn
e_n'&=&\frac{1}{512}\sqrt{(n-8)_8}\;,~~~
f_n'=\frac{(6n-11)}{192}\sqrt{(n-6)_6}\;,\nonumber\\
g_n'&=&\frac{1}{16}(2n^2-9n+7)\sqrt{(n-4)_4}\;,~~~
h_n'=-\frac{1}{64}(2n^3+93n^2-107n+66)\sqrt{(n-2)_2}\;,\nonumber\\
t_n'&=&-\frac{1}{64}(2n^3-123n^2-359n-300)\sqrt{n_2}\;,~~~
u_n'=\frac{1}{16}(2n^2+13n+18)\sqrt{n_4}\;,\nonumber\\
v_n'&=&\frac{(6n+17)}{192}\sqrt{n_6}\;,~~~
w_n'=\frac{1}{512}\sqrt{n_8}\;;
\label{xx3apa}
\eqn
\be
E_n^{(1)}=\frac{3}{4}(2n^2+2n+1)\Omega
\label{ap6a}
\ee
and

\be
E_n^{(2)}=-\frac{1}{8}(34n^3+51n^2+59n+21)\Omega\;.
\label{ap6b}
\ee
Writing Eq. (\ref{ap5}), in coordinate representation, in the form given
in Eq. (\ref{ec4}) we get for $G_n^{(1)}(\sqrt{\Omega}Q)$ and 
$G_n^{(2)}(\sqrt{\Omega}Q)$ respectively,

\be
G_{n}^{(1)}(\xi)=a_nH_{n-4}(\xi)+
b_{n}H_{n-2}(\xi)+c_{n}H_{n+2}(\xi)
+d_nH_{n+4}(\xi)\;
\label{xx1}
\ee
and

\bqn
G_{n}^{(2)}(\xi)&=&e_nH_{n-8}(\xi)+f_nH_{n-6}(\xi)+
g_nH_{n-4}(\xi)+h_nH_{n-2}(\xi)
\nonumber\\
& &
+t_nH_{n+2}(\xi)
+u_{n}H_{n+4}(\xi)+v_{n}H_{n+6}(\xi)+w_nH_{n+8}(\xi)\;,
\label{xx2}
\eqn
where $\xi=\sqrt{\Omega}Q$, 

\bqn
a_n&=&\frac{a_n'}{\sqrt{2^{n-4}(n-4)!}}\;,~~~
b_n=\frac{b_n'}{\sqrt{2^{n-2}(n-2)!}}\;,\nonumber\\
c_n&=&\frac{c_n'}{\sqrt{2^{n+2}(n+2)!}}\;,~~~
d_n=\frac{d_n'}{\sqrt{2^{n+4}(n+4)!}}
\label{xx3}
\eqn
and

\bqn
e_n&=&\frac{e_n'}{\sqrt{2^{n-8}(n-8)!}}\;,~~~
f_n=\frac{f_n'}{\sqrt{2^{n-6}(n-6)!}}\;,\nonumber\\
g_n&=&\frac{g_n'}{\sqrt{2^{n-4}(n-4)!}}\;,~~~
h_n=\frac{h_n'}{\sqrt{2^{n-2}(n-2)!}}\;,\nonumber\\
t_n&=&\frac{t_n'}{\sqrt{2^{n+2}(n+2)!}}\;,~~~
u_n=\frac{u_n'}{\sqrt{2^{n+4}(n+4)!}}\;,\nonumber\\
v_n&=&\frac{v_n'}{\sqrt{2^{n+6}(n+6)!}}\;,~~~
w_n=\frac{w_n'}{\sqrt{2^{n+8}(n+8)!}}\;.
\label{xx3a}
\eqn


\end{document}